\documentclass[11pt,aps,preprint,nofootinbib,superscriptaddress]{revtex4}
\usepackage[active]{srcltx}
\usepackage[utf8]{inputenc}
\usepackage{latexsym}
\usepackage{amsmath}
\usepackage{amsfonts}
\usepackage{graphicx}
\usepackage{slashed}
\usepackage{xcolor}
\usepackage{mathtools}

\begin{document}

\title{Bosonization of the Thirring Model in 2+1 dimensions}

\author{Rodrigo Corso B. Santos}
\email{rodrigocorso@uel.br}
\affiliation{Departamento de F\'isica, Universidade Estadual de Londrina, \\86057-970, Londrina, PR, Brasil}

\author{Pedro R. S. Gomes}
\email{pedrogomes@uel.br}
\affiliation{Departamento de F\'isica, Universidade Estadual de Londrina, \\86057-970, Londrina, PR, Brasil}

\author{Carlos A. Hernaski}
\email{carloshernaski@utfpr.edu.br}
\affiliation{Departamento de F\'isica, Universidade Estadual de Londrina, \\86057-970, Londrina, PR, Brasil}
\affiliation{Departamento de F\'isica, Universidade Tecnológica Federal do Paraná, 
85503-390, Pato Branco, PR, Brasil}

\begin{abstract}

In this work we provide a bosonized version of the Thirring model in 2+1 dimensions in the case of  single fermion species, where we do not have the benefit of large $N$ expansion. In this situation there are very few analytical methods to extract nonperturbative information. Meanwhile, nontrivial behavior is expected to take place precisely in this regime. To establish the bosonization of the Thirring model, we consider a deformation of a basic fermion-boson duality relation in 2+1 dimensions. The bosonized model interpolates between the ultraviolet and infrared regimes, passing several consistency checks and recovering the usual bosonization relation of the web of dualities in the infrared limit. In addition the duality predicts the existence of a nontrivial ultraviolet fixed point in the Thirring model.

\end{abstract}

\maketitle


\section{Introduction}
  
In the recent years a confluence of ideas in field theory and condensed matter has led to the discovery of an intricate set of dualities \cite{Karch,Witten,Murugan} thenceforth called web of dualities. This refers to a set of relations between 2+1-dimensional (3D) quantum field theories  valid at an infrared stable fixed point. In the heart of this web, one finds a relationship of the bosonization type, namely,
\begin{equation}
\bar{\psi}i \slashed{D}_A\psi -\frac{1}{2} \frac{AdA}{4\pi}~~~\Longleftrightarrow~~~ |D_a\phi|^2-\frac{\lambda_{\ast}}{4}|\phi|^4 + \frac{ada}{4\pi}+ \frac{Ada}{2\pi},
\label{0.2}
\end{equation}
from which many other dualities can be derived, including the original particle-vortex bosonic duality \cite{Peskin,Halperin} and its fermionic counterpart proposed in \cite{Son}. $D_a$ and $D_A$ are covariant derivatives acting on charge $+1$ fields. The left hand side corresponds to a free fermion coupled to a background field, $A$, while the right hand side  involves a complex scalar coupled to a compact dynamical gauge field with a Chern-Simons (CS) term and flux quantization over the sphere $\int_{S^2} da=2\pi\mathbb{Z}$, which is responsible for implementing the fermion-boson transmutation. The compact field couples to the external field through a $BF$ term $Ada$. In addition, this theory is taken at the Wilson-Fisher fixed point \cite{Wilson1}, so that both theories in (\ref{0.2}) are scale invariant.

The spirit of this work is to take advantage of this relation in order to extend the class of models related by duality. This is accomplished by considering deformations of (\ref{0.2}) and may be specially useful to explore nonperturbative regimes, where usually we do not have much analytical methods at our disposal. A possible deformation of the duality (\ref{0.2}) is to include mass operators in both sides \cite{Tong,Witten}. In this case, the duality assumes the form
\begin{equation}
\bar{\psi}i \slashed{D}_A\psi-M\bar{\psi}\psi-\frac{1}{2} \frac{AdA}{4\pi}~~~\Longleftrightarrow~~~ |D_a\phi|^2-m^2|\phi|^2-\frac{\lambda}{4}|\phi|^4 + \frac{ada}{4\pi}+ \frac{Ada}{2\pi}.\label{0.21}
\end{equation}
Of course the parameters of the two theories must be somehow connected. The above relation has been verified in the strict limit of infinite masses, reproducing correctly the Hall conductivity (Chern-Simons coefficient) in both sides \cite{Tong,Witten}. A more precise relation between the masses was suggested in \cite{Tong}, motivated by an analogy with large $N$ studies \cite{Jain,Ari}. The idea is to introduce an auxiliary field, $\sigma$, which allows us to write the interaction $-\frac{\lambda}{4}|\phi|^4$ as $-\sigma|\phi|^2+\frac{1}{\lambda}\sigma^2$. The map between operators is then established in terms of the auxiliary field, according to $\bar{\psi}\psi \sim-\sigma $. Including a mass term for the fermion, $-M\bar{\psi}\psi$, is equivalent to add on the bosonic side the operator $M\sigma$, which in turn implies the identification between the masses in (\ref{0.21}) as $m^2 \sim -\lambda M$. As we shall see, the consistency of this identification emerges as a by-product of the analysis in the present work. 

It is worth emphasizing that the infrared duality that follows from (\ref{0.21}) can be explicitly derived simply by computing the Hall conductivity in both sides. Both models describe two gapped phases depending on the sign of $M$ ($m^2$) on the fermionic (bosonic) side. If the phase transition on the bosonic side is of second order, one can conjecture that the two models share the same critical point described by the duality (\ref{0.2}) \cite{Witten}. By assuming this conjecture we have an exact relation between conformal theories valid, therefore, for arbitrary energies. By adding operators depending on the background field in both sides of the duality and promoting them to dynamical fields leads to new dualities. Making this procedure in (\ref{0.21}) we get new dualities that must be true in the infrared. However, if this is done in (\ref{0.2}) we get new conjectured dualities valid for arbitrary energy scales. As we shall see, a particularly interesting operator that fits in this perspective is the Thirring interaction, i.e., a four-fermion interaction of the form $(\bar{\psi}\gamma^{\mu}\psi)^2$, which is extremely important in several contexts.

Indeed, the 3D Thirring model (more generally, models with four-fermion interactions) has served over the years as valuable prototype to examine a number of methodological questions, such as large $N$ renormalizability \cite{Warr,Gomes,Park}, realization of the Weinberg's asymptotic safety scenario \cite{Braun}, dynamical symmetry breaking \cite{Rosenstein}, lattice simulations \cite{Christofi,Lucini,Debbio,Wipf,Hands3}, and the relationship to other models like $\text{QED}_3$ \cite{Hands1,Itoh}. Furthermore, it is relevant for condensed matter systems of great interest, as in the case of high-$T_c$ superconductors \cite{Herbut,Franz} and also in  the description of low-energy excitations of materials like graphene \cite{Strouthos,Roy}.

In this work we investigate the 3D Thirring model with a single fermion field $(N=1)$, in the light of the web of dualities. A dual description of the Thirring model in 3D offers in principle the possibility to explore strong coupling regimes, which are difficult to be accessed by analytical methods like large $N$ expansion\footnote{In this context, it is worth to mention an interesting recent approach to deal with fermions in 3D in terms of quantum wires \cite{Mross1,Mross2,Hernaski}, where one spatial dimension is discretized so that the 3D problem is transformed into a set of 2D ones, and all the machinery of 2D bosonization can be used.}. By carrying out simple manipulations of the partition functions we are able to obtain a bosonized version of the Thirring model from the critical duality (\ref{0.2}) in the presence of an external field.

To support the proposed duality we analyse some accessible limits of the relation. Firstly, we consider the strict limit of infinite mass, where we recover the Fradkin-Schaposnik map relating the fermionic theory with the Maxwell-Chern-Simons model \cite{Fradkin}. Then we consider the limit of large but finite mass, where we perform an expansion in the inverse of mass, matching both sides to the first leading terms. In particular, this matching confirms the relation $m^2\sim-\lambda M$ suggested in \cite{Tong}.

This work is organized as follows. In Sec. \ref{S2} we review some useful properties of the Thirring model in 3D. In Sec. \ref{S3}, we discuss how to obtain the bosonized version of the Thirring model starting from the critical duality. In Sec. \ref{S4}, some checks of the proposed duality are performed. We conclude in Sec. \ref{S5} with a brief summary and additional comments.

 
\section{3D Thirring Model}\label{S2}

We start by discussing some features of the Thirring model in 2+1 dimensions given by the action
\begin{equation}
S=\int d^3x\left[ i\bar{\psi}\slashed\partial \psi -M \bar{\psi}\psi -\frac{g}{2} (\bar{\psi}\gamma^{\mu}\psi)^2\right].
\label{1.1}
\end{equation}
We are considering the two-component irreducible representation for the Dirac spinors. Most of literature about the Thirring model consider the reducible four-dimensional representation, where generally there is no Chern-Simons generation in the effective action of fermions coupled to gauge fields. We insist in the case of the irreducible representation since this is precisely the case that takes place in the web of dualities, where the Chern-Simons term has a fundamental role.
The Thirring model involves two dimensionfull parameters: the mass, $[M]=1$, and the coupling constant, $[g]=-1$. As the coupling constant has negative mass dimension, the Thirring operator is irrelevant at weak coupling. On the other hand, large $N$ analysis has shown that the model is renormalizable in this framework \cite{Gomes,Park} suggesting it has a nonperturbative UV fixed point.

The essential ingredient in the fermion-boson dualities in 2+1 dimensions is the particle-vortex nature of the mapping between objects of the dual theories. In other words, a local field that creates a particle excitation in one theory corresponds to a monopole operator in its dual \cite{Karch,Witten}. A different kind of bosonization duality for the Thirring model has been discovered by Fradkin and Schaposnik in \cite{Fradkin}. In that work the authors show that the strict large mass limit of the Thirring model is equivalent to the Maxwell-Chern-Simons (MCS) theory. More precisely, they consider the energy regime $E\sim\frac{1}{g}\ll |M|$. Since no single fermion state can be excited in the large mass limit, the mapping actually shows that the bound-state sector\footnote{In the large $N$ expansion, the bound-state condition can be determined exactly by examining the pole structure of the auxiliary vector field $A_{\mu}$ arising when we write the Thirring interaction as $-\frac{g}{2} (\bar{\psi}\gamma^{\mu}\psi)^2= \frac{A_{\mu}^2}{2g}-A_{\mu}\bar{\psi}\gamma^{\mu}\psi$, which exhibits a pole satisfying $k^2<4M^2$ when $g>-\frac{4\pi}{|M|}$ \cite{Gomes}. Thus, in the large mass limit, for any positive coupling constant $g$ (corresponding to attractive interactions) there is bound-states formation.} of the Thirring model can be described by a bosonic gauge field governed by the MCS dynamics. Therefore the large mass limit of the Thirring model is described by a purely bosonic gauge theory
\begin{align}
i\bar{\psi}\slashed\partial \psi -M \bar{\psi}\psi -\frac{g}{2} (\bar{\psi}\gamma^{\mu}\psi)^2 ~~~\overset{|M|\rightarrow\infty}{\Longleftrightarrow}~~~-\frac{g}{64\pi^2}f_{b}^{2}-\text{sign}(M)\frac{ bdb}{8\pi},
\label{2.6}
\end{align}
where the sign of the fermion mass determines the sign of the CS coefficient in the bosonic description. In \cite{Banerjee} a bosonized version of the Thirring model is obtained that is valid to all orders in an expansion in the inverse of mass. Since no spin and statistical transmutation are involved in the Fradkin-Schaposnik mapping, the compact character of the emergent field, and consequently monopole configurations, can be ignored. We will show that this mapping can be recovered from the large mass limit of a more complete bosonized version of the Thirring model for finite mass. It is important to emphasize that the relation (\ref{2.6}) is valid even in the strong coupling limit and thus provides an useful consistency check for the duality that we will discuss in the next sections.


\section{Thirring Model from the Critical Duality} \label{S3}

In the light of the previous discussions, we will see now how to induce the Thirring operator in the critical duality (\ref{0.2}) by adding operators involving exclusively the external field. In this way, we propose the following duality relation:
\begin{eqnarray}
&&\bar{\psi}i \slashed{D}_A\psi-\frac{g}{2}(\bar{\psi}\gamma^{\mu}\psi)^2 ~~~\Longleftrightarrow~~~|D_{a}\phi|^2-\frac{\lambda}{4}|\phi|^4+\frac{1}{4\pi}ada\nonumber\\
&-&\frac{1}{2\pi}cdc-\frac{g}{16\pi^2}f_{a+c}^2+\frac{1}{2\pi}\left(a+c\right)dA,
\label{3.1}
\end{eqnarray}
where $a$ is an emergent compact gauge field satisfying the Dirac flux quantization condition
\begin{equation}
\int_{S^2} da=2\pi\mathbb{Z},
\label{3.21}
\end{equation}
with $S^2$ being a compactification of the space slices $\mathbb{R}^2$. The field $c$ is an independent gauge field with no flux quantization condition. Some comments are in order. Firstly, the flux quantization condition ensures that the partition function for the bosonic model is gauge invariant including large gauge transformations of $a$. In second place, naturally the above relation contains (\ref{0.2}). Indeed, by flowing to low energies the Thirring operator becomes irrelevant, and we recover the original duality (taking $g\rightarrow 0$ we can directly integrate out the gauge field $c$ in the bosonic theory to obtain (\ref{0.2})). In the next section we will see that, in regimes of energy where the Thirring operator is important, the above duality passes some consistency checks and predicts an interesting behavior in the UV limit. In the following we shall discuss how to obtain the above relation.

To motivate the relation (\ref{3.1}), we start with the critical duality (\ref{0.2}) and introduce in both sides the
term $\frac{1}{2g}(A-B)^2$, with $B$ being a new external field. This sets out an energy scale, $g^{-1}$, in the duality that was previously scale invariant. Then, promoting $A$ to a dynamical field and renaming the fields as $A\rightarrow eb$ and $B\rightarrow A$, we obtain
\begin{eqnarray}
&&\bar{\psi}i (\slashed{\partial}-ie\slashed{b})\psi+\frac{1}{2g}(eb-A)^2~~~\Longleftrightarrow~~~ |D_a\phi|^2-\frac{\lambda}{4}|\phi|^4+\frac{1}{4\pi}ada\nonumber\\\nonumber\\&+&\frac{e}{2\pi}bda+\frac{e^2}{8\pi}bdb+\frac{1}{2g}(eb-A)^2.
\label{3.3}
\end{eqnarray}
At this point $b$ is a non-compact field. Notice also that in promoting $A$ to a dynamical field we have assigned a corresponding coupling constant $e$, which is useful to properly analyse specific regimes of energy. We see that the left hand side produces the Thirring interaction after integrating out $b$. Actually, we first make the shift $eb\rightarrow eb+A$, and then integrate out this redefined field producing the Thirring interaction with a coupling constant $g$. In order to obtain the proposed duality, the right hand side still needs some attention. Let us consider the part of the Lagrangian involving the $b$ field:
\begin{equation}
\mathcal{L}[a,b,A]\equiv \frac{e}{2\pi}bda+\frac{e^2}{8\pi}bdb+\frac{1}{2g}(eb-A)^2. \label{3.5}
\end{equation}
In the spirit of the connection between Self-Dual and MCS models \cite{Deser}, we define the interpolating Lagrangian
\begin{equation}
\mathcal{L}_I=-\frac{1}{8\pi}cdc+\frac{e}{4\pi}cdb+\frac{1}{2g}\left(eb-A\right)^{2}+\frac{e}{2\pi}bda,\label{3.6}
\end{equation}
where $c$ is a new emergent gauge field and similarly to $b$ is also non-compact. Integrating over $c$ gives $c=eb$. Plugging this relation back into the Lagrangian (\ref{3.6}) gives (\ref{3.5}). Therefore, (\ref{3.5}) and (\ref{3.6}) are equivalent. Alternatively, we can integrate out the field $b$ to get $eb=A-\frac{g}{4\pi}\left(dc+2da\right)$. Plugging this back into the Lagrangian (\ref{3.6}) and making the field redefinition $ c \rightarrow 2 \tilde{c} $, we obtain
\begin{align}
\mathcal{L}=-\frac{1}{2\pi}\tilde{c}d\tilde{c}-\frac{g}{16\pi^{2}}f_{a+\tilde{c}}^{2}+\frac{1}{2\pi}\left(a+\tilde{c}\right)dA. \label{3.8}
\end{align}
Therefore, we can see (\ref{3.6}) as an interpolating Lagrangian between (\ref{3.5}) and (\ref{3.8}). Replacing (\ref{3.5}) by (\ref{3.8}) in (\ref{3.3}) and renaming $\tilde{c}\rightarrow c$, we obtain the right hand side of (\ref{3.1}).

Now, we analyse the field content of (\ref{3.1}) and determine the fermion operator in terms of the bosonic field and the monopole operator (bosonization rule). The charges of the monopole operator $\mathcal{M}_a$ can be read out from the respective CS and BF coefficients involving the field $a$. This is summarized in Table \ref{table1}. The fermion is identified as the operator that is uncharged by the emergent gauge fields $a$ and $c$, being charged only by the electromagnetic field $A$:
\begin{equation}
\psi ~~~\Longleftrightarrow~~~ \phi^{\dagger} \mathcal{M}_a.
\end{equation}
As the combination of fields $a+c$ is coupled to the electromagnetic field $A$, we can read out the corresponding current $\bar{\psi}\gamma^{\mu}\psi \Leftrightarrow\frac{1}{2\pi}\epsilon^{\mu\nu\rho}\partial_{\nu}(a_{\rho}+c_{\rho})$. 
\begin{table}[htb]
\centering
\large
\begin{tabular}{|c|c|c|}
\hline
 &
$U(1)_a$
&  $U(1)_A$
\\
\hline \hline
$\mathcal{M}_a$ & 1 & 1 \\ \hline
$\phi$ & 1 &  0\\  \hline
\end{tabular}
\caption{Charge of the monopole operators and of the scalar field. }
\label{table1}
\end{table}

An immediate consequence of the duality (\ref{3.1}) is that it points out to the existence of a nontrivial fixed point in this the UV regime. Indeed, considering the bosonic dual model, a simple dimensional analysis shows that, since $\lambda$ and $g^{-1}$  have positive mass dimension, the operators $|\phi|^4$, $ada$, and $cdc$ are negligible at the UV implying that the bosonic model has a trivial UV fixed point. This, in turn, must correspond to a strongly coupled fixed point on the fermionic side, according to the relation
\begin{equation}
\bar{\psi}i \slashed{\partial}\psi-\frac{g_\ast}{2}(\bar{\psi}\gamma^{\mu}\psi)^2 ~~~\Longleftrightarrow~~~|\partial\phi|^2-\frac{g_\ast}{16\pi^2}f_{a+c}^2.
\label{4.14}
\end{equation}
Therefore, the duality predicts an UV nontrivial fixed point for the Thirring model. Recent studies employing nonperturbative methods also provide strong evidence for the existence of nontrivial fixed points in several four-fermion theories in 2+1 dimensions \cite{Hands,Lenz,Gies,Gies2,Dabelow}. However, a more quantitative comparison with our result is not available since these previous studies consider the four-dimensional representation for the Dirac spinors, which induces a couple of independent four-fermion interactions, contrarily to the case of the irreducible representation.


\section{Testing the Duality}\label{S4}

In order to make some quantitative tests of the duality, it is convenient to extend it to the massive case. With this purpose, we assume that the massive form of relation (\ref{0.21}) remains valid in the present case, so that we consider the following relation
\begin{eqnarray}
&&\bar{\psi}i \slashed{D}_A\psi-M\bar{\psi}\psi-\frac{g}{2}(\bar{\psi}\gamma^{\mu}\psi)^2 ~~~\Longleftrightarrow~~~|D_{a-c}\phi|^2-m^2|\phi|^2-\frac{\lambda}{4}|\phi|^4+\frac{1}{4\pi}ada\nonumber\\&-&\frac{1}{4\pi}cdc-\frac{1}{2\pi}adc-\frac{g}{16\pi^2}f_a^2+\frac{1}{2\pi}adA.
\label{3.1massive}
\end{eqnarray}
For convenience we have made the shift $a\rightarrow a-c$, which leads to a more direct process of integration over fields. Notice that in this basis, both scalar field and the monopole operator $\mathcal{M}_a$ are charged under $c$. Nevertheless, the combination $\phi^{\dagger}\mathcal{M}_a$ remains uncharged.

Now, we consider the energy regimes where the Thirring operator becomes more important. Actually, the duality can be quantitatively tested in the limit of large but finite mass, i.e., for energies $E\ll |M|$, independent of the value of $g$. In this way, the relations that emerge in this limit remain valid even for the strong-coupling regime. We emphasize that, as the scalar excitations are suppressed in this regime, no spin and statistical transmutation occur and therefore we can neglect monopole configurations.

On the fermionic side the calculation is straightforward. Introducing the mass term in (\ref{3.3}) and integrating out the fermion field, we get
\begin{eqnarray}
\frac{e^2\text{sgn}(M)}{8\pi}bdb-\frac{e^2}{48\pi|M|}f^2_b+\frac{1}{2g}\left(eb-A\right)^2+O\left(1/M^2\right).\label{4}
\end{eqnarray}
We consider the alternative Lagrangian
\begin{eqnarray}
-\frac{\text{sgn}(M)}{8\pi}cdc+\frac{e}{4\pi}cdb-\frac{1}{48\pi|M|}f^2_c+\frac{1}{2g}\left(eb-A\right)^2.\label{64}
\end{eqnarray}
Integrating $c$ out, we get
\begin{eqnarray}
c^{\mu}&=&\text{sgn}(M)eb^{\mu}-\frac{e^2}{6|M|}\epsilon^{\mu\nu\sigma} f_{\nu\sigma}^b+O\left(1/M^2\right),\label{65}
\end{eqnarray}
where we have used (\ref{65}) recursively to write the Maxwell term for the field $c$ in terms of the field $b$ (the difference being of $O\left(1/M^2\right))$. Plugging this back into (\ref{64}), we obtain
\begin{equation}
\frac{\text{sgn}(M)e^2}{8\pi}bdb-\frac{e^2}{48\pi|M|}f^2_b+\frac{1}{2g}\left(eb-A\right)^2+O\left(1/M^2\right),
\end{equation}
which is equivalent to (\ref{4}) up to $O\left(1/M^2\right)$ terms.

Now, integrating out the $b$ field in (\ref{64}), we have $eb=A-\frac{g}{4\pi}dc$. Plugging this back into (\ref{64}), we finally get
\begin{eqnarray}
-\frac{\text{sgn}(M)}{8\pi}cdc+\frac{1}{4\pi}cdA-\left(\frac{g}{64\pi^2}+\frac{1}{48\pi|M|}\right)f^2_c.
\label{70}
\end{eqnarray}

On the bosonic side things are slightly more subtle. The strategy is similar to the fermionic case, in the sense that we will integrate out the scalar field in the large mass limit. The resulting effective action is then organized in a loop expansion (with internal lines involving only the scalar field) with parameter $\kappa\equiv \frac{1}{(4\pi)^{\frac32}}\frac{\lambda}{|m|}$. The difference, however, is that we have to consider the two phases of the model separately. 

Let us examine firstly the Higgs phase, where $m^2<0$, so that the combination $(a-c)$ acquires a mass. Expanding around the new vacuum $v=\left<\phi\right>=\sqrt{-\frac{2m^{2}}{\lambda}}$ on the RHS of (\ref{3.3}) and integrating out the massive scalar mode, we obtain
\begin{equation}
-\frac{2m^2}{\lambda}\left(1+O(\kappa)\right)(a-c)^2-\frac{1}{4\pi}cdc+\frac{1}{4\pi}ada-\frac{1}{2\pi}adc-\frac{g}{\left(4\pi\right)^2}f_a^2+\frac{1}{2\pi}Ada.
\label{76}
\end{equation}
Integrating $c$ out, we have $c^{\mu}=a^{\mu}-\frac{\lambda}{4 \pi m^{2}}\epsilon^{\mu \nu \sigma} \partial_{\nu} a_{\sigma}+ O\left(\frac{\lambda^{2}}{m^{4}}\right)$, where we have used the equation of motion for $c$ recursively to replace $dc$ by $da$. Using this relation, we finally get
\begin{eqnarray}
&&-\frac{1}{2\pi}ada-\left(-\frac{\lambda}{16\pi^2m^2}+\frac{g}{16\pi^2}\right)f_a^2+\frac{1}{2\pi}Ada.
\label{83}
\end{eqnarray}
Comparing the Chern-Simons coefficient  of (\ref{83}) with (\ref{70}) (after rescaling $ a \rightarrow a/2 $), we see that this phase corresponds to the case with $M>0$ on the fermionic side. Furthermore, matching the Maxwell term gives the relation 
\begin{equation}
m^2=-\frac{3}{4\pi}|M|\lambda.
\label{4.9}
\end{equation}
This result is consistent with the mechanism proposed in \cite{Tong} to deform the critical duality by the addition of mass terms in both sides, which leads to a relation of the type $m^2\sim -M\lambda$.

In the symmetric phase, by integrating out the scalar field gives
\begin{eqnarray}
-\frac{1}{96\pi|m|}\left(1+O(\kappa)\right)f^2_{a-c}-\frac{1}{4\pi}cdc+\frac{1}{4\pi}ada-\frac{1}{2\pi}adc-\frac{g}{\left(4\pi\right)^2}f_a^2+\frac{1}{2\pi}Ada.\label{72}
\end{eqnarray}
Integrating $c$ out, we have $c=-a+\frac{1}{12|m|}da+O(1/m^2)$. Plugging this back into (\ref{72}), we obtain
\begin{eqnarray}
&&-\left(\frac{g}{16\pi^2}+\frac{1}{24\pi|m|}\right)f^2_a+\frac{1}{2\pi}ada+\frac{1}{2\pi}Ada,
\label{74}
\end{eqnarray}
which can be compared to (\ref{70}) after rescaling $a\rightarrow a/2$. We see that this corresponds to the fermionic phase with $M<0$. Therefore we can write an identification between masses in an unified way,  
\begin{equation}
m^2=-\frac{3}{4\pi}M\lambda,
\label{4.9a}
\end{equation}
which is valid in the two phases, i.e.,  $M>0 \Rightarrow m^2<0$ (Higgs phase) and $M<0 \Rightarrow m^2>0$ (symmetric phase). Comparison of the Maxwell term fixes the value of $\frac{\lambda}{|m|}$,
\begin{equation}
\frac{\lambda}{|m|}=\frac{2\pi}{3}.
\label{4.12}
\end{equation}
This is the specific point at which the duality is valid in the regime of energy $E\ll |m|$. The fact that the coupling constant $\lambda$ is not a free parameter is a consistency requirement, since the fermionic side does not contain any dependence on $\lambda$. This also happens in the original duality, which holds at the Wilson-Fisher fixed point. Furthermore, we see that the  value (\ref{4.12}) leads to a properly perturbative parameter $\kappa$, 
\begin{equation}
\kappa\equiv \frac{1}{(4\pi)^{\frac32}}\frac{\lambda}{|m|}=\frac{1}{12\pi^{\frac12}}\sim 0.047,
\end{equation}
consistent with the initial assumption that $\kappa$ is small.

It is interesting to notice that the matching of the Maxwell terms is automatic for the Thirring coupling $g$. This indicates that our calculation also gives a matching for the original duality (\ref{0.21}) to the order of $1/M$ if we turn off $g$. On the other hand, in the strict limit $|M|\rightarrow\infty$, it gives the mapping
\begin{eqnarray}
i\bar{\psi}\slashed{D}_A \psi -M \bar{\psi}\psi -\frac{g}{2} (\bar{\psi}\gamma^{\mu}\psi)^2~~~\overset{|M|\rightarrow\infty}{\Longleftrightarrow}~~~-\frac{\text{sgn}(M)}{2\pi}ada+\frac{1}{2\pi}Ada-\frac{g}{16\pi^2}f^2_a,\label{4.13}
\end{eqnarray}
which, after the rescaling $a\rightarrow a/2$, is precisely the Fradkin-Schaposnik mapping \cite{Fradkin} shown in (\ref{2.6}). There is a massive excitation appearing on the right hand side with mass $\sim{1}/{g^2}$. This corresponds to a fermion-antifermion bound-state on the left hand side. In the UV limit, where the Thirring model is strongly coupled, $g\rightarrow\infty$, the bound state becomes massless.


\section{Conclusions}\label{S5}

We have provided a bosonization of the Thirring model in 2+1 dimensions for the case of a single fermion specie with arbitrary mass. Therefore it extends previous works on the Thirring model that relied mostly on large $N$ techniques or large mass limit. This was obtained from the bosonization duality of a Dirac fermion coupled to a background gauge field, which lies in the heart of the web of dualities. By carrying out manipulations of the corresponding partition functions we were able to produce the Thirring model on one side and the corresponding bosonic theory on the other. A number of consistency checks were performed, including the infinite mass limit where we recover the known Fradkin-Schaposnik map, placing it in the context of the web of dualities. We have also explicitly shown the validity of the duality for large but finite mass, which encompasses automatically the original duality. As a by-product of our analysis, we confirmed the relation implied in \cite{Tong} connecting the mass parameters of the models and the coupling constant of the quartic bosonic interaction. 

The duality we have proposed for the Thirring model involves on the bosonic side a combination of gauge fields, where the four-fermion coupling constant is directly tied to the Maxwell term. Thus the highly nontrivial task of analyzing the strong coupling properties of the Thirring model can become manageable using its dual version since it corresponds to a quadratic term. Indeed, a simple dimensional analysis shows that the bosonic model has a trivial fixed point in the UV. The duality relation then implies that this should correspond to a nontrivial UV fixed point in the Thirring model. This is an important physical information that is extremely hard to be extracted when we do not have benefits of the large $N$ or large mass expansions.

In the above context, it is interesting to make further contact with the work \cite{Tong}, which suggests that the UV fixed point of the Gross-Neveu model is described in terms of a free scalar theory. Up to a decoupled Maxwell term, this is the type of theory that emerges in the UV limit of our duality for the Thirring model. We believe that this is a reflection of the fact that the Thirring interaction in 3D with a single spinor in the irreducible representation is equivalent to the Gross-Neveu interaction. In this sense, the UV limit of our duality is in compliance with that one proposed in \cite{Tong}.

We conclude by saying that further checks of the duality proposed in this work are currently under investigation and shall be reported elsewhere. Nevertheless, we expect that discussions in this work could be helpful in the understanding of strongly interacting four-fermion theories in 2+1 dimensions.


\section{Acknowledgments}

We are grateful to Holger Gies for the useful correspondence concerning the existence of the UV nontrivial fixed point in the Thirring model. We acknowledge the financial support of Brazilian agencies CAPES and CNPq.


\end{document}